# Extracting impedance changes from a frequency multiplexed signal during neural activity in sciatic nerve of rat: preliminary study in-vitro

J. Hope[1,2], K. Aristovich[3], C. A. R. Chapman[3], A. Volschenk[4], F. Vanholsbeeck[2,5], A. McDaid[1]

[1] Department of Mechanical Engineering, The University of Auckland, NZ
[2] Dodd Walls Centre for photonic and Quantum Technologies, NZ
[3] The Department of Medical Physics and Biomedical Engineering, University College London, UK
[4] The School of Biological Sciences, The University of Auckland, NZ
[5] The Department of Physics, The University of Auckland, NZ

**Abstract**

**Objective:** Establish suitable frequency spacing and demodulation steps to use when extracting impedance changes from frequency division multiplexed (FDM) carrier signals in peripheral nerve.

**Approach:** Experiments were performed *in-vitro* on cadavers immediately following euthanasia. Neural activity was evoked via stimulation of nerves in the hind paw, while carrier signals were injected, and recordings obtained, with a dual ring nerve cuff implanted on the sciatic nerve. Frequency analysis of recorded compound action potentials (CAPs) and extracted impedance changes, with the latter obtained using established demodulation methods, were used to determine suitable frequency spacing of carrier signals, and bandpass filter (BPF) bandwidth and order, for a frequency multiplexed signal.

**Main results:** CAPs and impedance changes were dominant in the frequency band 200 – 500 Hz and 100 – 200 Hz, respectively. A Tukey window was introduced to remove ringing from Gibbs phenomena. A +/- 750 Hz BPF bandwidth was selected to encompass 99.99% of the frequency power of the impedance change. Modelling predicted a minimum BPF order of 16 for 2 kHz spacing, and 10 for 4 kHz spacing, were required to avoid ringing from the neighbouring carrier signal, while FDM experiments verified BPF orders of 12 and 8, respectively, were required. With a notch filter centred on the neighbouring signal, a BPF order of at least 6 or 4 was required for 2 and 4 kHz, respectively.

**Significance:** The results establish drive frequency spacing and demodulation settings for use in FDM electrical impedance tomography (EIT) experiments, as well as a framework for their selection, and, for the first time, demonstrates the viability of FDM-EIT of neural activity on peripheral nerve, which will be a central aspect of future real-time neural-EIT systems and EIT-based neural prosthetics interfaces.

1. Introduction

Functional monitoring of neural activity in peripheral nerves is a sought after goal in neural prosthetics. Such monitoring should be capable of identifying multiple concurrent neural signals in real time, and with a sub-millisecond temporal resolution. While intra-neural electroneurography (ENG) technologies, such as the UTAH slanted electrode array, can resolve multiple concurrent sites of activity in real-time, and with spatial and temporal resolutions suitable for neural prosthetics [3], they have a high risk of medical complications due to the crush and cut type injury during device insertion, and persistent inflammation from stiffness mismatch between electrodes and neural tissues [4, 5]. Extra-neural ENG-technologies, on the other hand, are considered safe for chronic implantation and are compatible with real-time applications [6], but have not been shown to spatially resolve more than one [7-9] or two [10] concurrently active sites of neural activity, the former with nerve cuff and the latter with in flat interface nerve electrodes.

Electrical Impedance Tomography (EIT), an imaging modality that resolves the impedance distribution within a sample, is compatible with nerve cuff [2] and is sensitive to neural activity due to the transient change in impedance associated with neuronal depolarisation [11]. In EIT, a sinusoidal drive current is passed between two drive electrodes and the resultant boundary voltages are recorded using several other measurement electrodes, producing a single sub-set of measurement data. Repetition of this process from several different 'angles', or active drive electrode pairs, is required to produce a full set of measurement data for each temporal increment. The benefits of neural-EIT over inverse source analysis of passive ENG recordings are superior spatial resolution [2], and the potential, as yet unconfirmed by *in-vivo* or *in-vitro* experiments, to resolve multiple concurrent sites of neural activity [12]. The draw backs of EIT, are the lower signal to noise ratio (SNR) [2], additional hardware [13], and an inferior temporal resolution [14-16]. A multi-frequency EIT system uses frequency division multiplexing (FDM), whereby two or more drive currents at different carrier frequencies are simultaneously active, to improve temporal resolution [17]. Similarly, phase-division multiplexing (PDM), whereby two drive currents at the same frequency but 90 degree offset in phase are simultaneously active, can also increase temporal resolution, and is compatible with frequency multiplexing [18].

Neural-EIT experiments have been performed under controlled conditions to image cortical activity in 3D [14] and extract transient impedance changes [19] during sensory stimuli in cerebral cortex of rat *in-vivo*; to image cortical activity in 2D [15, 20] and extract impedance changes [21] during ictal and interictal activity from induced epileptic seizures in rat *in-vivo*; to image fascicle level activity [2, 16] or extract transient impedance changes [22] in sciatic nerve of rat *in-vivo*; and, to extract transient impedance changes from non-mammalian nerve *ex-vivo* [23]. In each study, transient impedance changes were extracted using the following steps, in order: time-domain windowing of the boundary voltage measurements, bandpass filter (BPF) with bandwidth centred on the drive current frequency, and then either the Hilbert transform (HT) [15, 19, 21-23] or an un-noted signal envelope function [2, 14, 16, 20]. Furthermore, with the exception of [22] where the bandwidth was +/-50 Hz, the bandwidth of the BPF reflected the frequency components of the underlying transient impedance changes from neural activity, which, as one would expect from the temporal duration of the associated neural activity, are lower in unmyelinated axons of non-mammalian nerves: +/-125 Hz [23], and cerebral cortex: +/-125 to +/- 500 Hz [14, 15, 19, 20], than in myelinated axons: +/- 2 to +/- 3 kHz [2, 16]. While most of these neural EIT studies investigated drive currents at more than one frequency [2, 14-16, 19-22], only one, a preliminary study on induced epileptic seizures in rat *in-vivo* at 2.2 and 2.6 kHz, implemented FDM-EIT [20]. Investigations into PDM-EIT of neural activity have, to date, only been performed on resistor phantom [18].

*1.1. Theory*

In EIT of neural activity, boundary voltage measurements $y(t)$, expressed as a function of time $t$, are comprised of a contribution from compound action potential (CAP), and a contribution from the EIT drive current, which we may term the 'carrier signal', equation 1:

$$y(t) = v_c \cos(\omega_0 t) z(t) + v_{AP}(t) + v_s(t) + v_n(t) \tag{1}$$

where $v_c \cos(\omega_0 t)$ is the carrier signal with amplitude $v_c$ and angular frequency $\omega_0$; $z(t)$ is the tissue impedance change, which is normalised against the basal value of the sample in the inactive state, and which amplitude-modulates the carrier signal as the tissue undergoes transient changes in impedance during neuronal depolarisation, Fig 1a and b; $v_{AP}(t)$ is the CAP, Fig 1a; $v_s(t)$ is the voltage artefact from the stimulation pulse, or stimulus artefact, used to artificially excite neural tissue in bench-top experiments; and $v_n(t)$ is noise. The Fourier transform of $y(t)$ is given by:

$$Y(\omega) = \tfrac{1}{2} v_c \left[ Z(\omega - \omega_0) + Z(\omega + \omega_0) \right] + V_{AP}(\omega) + V_s(\omega) + V_s(\omega) \tag{2}$$

where $\omega$ is angular frequency, and capital letters $Y, Z$ and $V$ indicate the Fourier transform of the corresponding function with lower case letters. In the frequency domain, equation 2, it is clear that suitable selection of $\omega_0$ can be used to separate the CAP from the carrier signal, Fig 1b, and a BPF centred around $\omega_0$ and with bandwidth which encapsulates the main frequency components of $Z(\omega)$ will extract the impedance change.

The limited operating frequency range in peripheral nerve, estimated to be 2 – 20 kHz in [1], makes it desirable in an FDM system to place carrier signals as closely together as possible in the frequency domain, while still maintaining the integrity of the amplitude and shape of the demodulated impedance change due to its importance in EIT reconstruction.

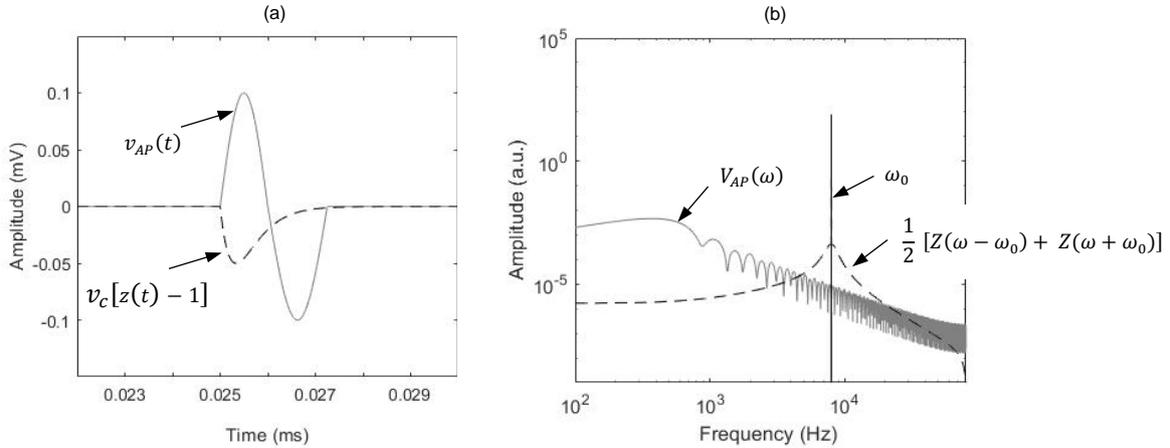

Figure 1: (a) Example CAP $v_{AP}(t)$ modelled using a positive half period of a 500 Hz sine wave appended to the negative half period of a 400 Hz sine wave to approximate results in [1], and example impedance change $v_c[z(t) - 1]$, shifted to a basal value of zero for comparison to the CAP, modelled using $z(t) = 1 - ate^{(bt)}$ with $a$ and $b$ selected as 4.08 and $-3 \times 10^3$, respectively, to produce an amplitude of $-5 \times 10^{-4}$ (-0.05% [2] of a 100mV carrier signal $v_c$), and duration of 1.5 ms. (b) Frequency components of the CAP, impedance change, and an 8 kHz carrier signal, plotted on log-log scale due to the small amplitude of the two former relative to the latter.

*1.2. Current Study*

In the current study, we extract transient impedance changes from FDM drive currents during, for the first time, evoked neural activity in sciatic nerve of rat *in-vitro*. To achieve this we apply a tapered window followed by the BPF-HT demodulation method established in [15, 19, 21-23], within which we design the BPF using frequency component analysis of the neural activity's CAP and impedance components. The results provide insight into suitable drive frequencies in FDM-EIT experiments, as well as a framework for their selection, and demonstrates the viability of FDM-EIT of neural activity on peripheral nerve, which will be a central aspect of future real-time neural-EIT systems and EIT-based neural prosthetics interfaces.

## 2. Materials and Methods

### 2.1. Tissue preparation and handling

Animal procedures were approved by the University of Auckland Animal Ethics Advisory Committee. A total of 5 animals were used; all Sprague-Dawley breed rats, male, and weighing 500 – 650 grams. Experiments were performed *in-vitro* on cadavers immediately following euthanasia using Carbon Dioxide followed by cervical dislocation, and performed as part of routine colony population management. On each animal, the left hind leg was de-gloved of skin then an incision was made down the posterior side between the Gracilis and Biceps Femoris muscles, exposing the sciatic nerve and proximal sections of its distal branches within the intermuscular cavity. Tissue of the Gracilis muscle was cut away to improve access to the sciatic nerve and facilitate implant of the nerve cuff. During the course of experiments, 0.01 M phosphate buffered solution (PBS), oxygenated with Carbogen 5 gas (95% $O_2$ / 5% $CO_2$) and heated to 38 ºC, was administered to exposed surfaces of the nerve branches and muscle to prolong tissue life.

The three main nerve branches which bifurcate from the sciatic nerve – the tibial, peroneal and sural branches – innervate the hind paw [24]. Artificially evoked neural activity in the sciatic nerve was, therefore, achieved by administering a stimulation pulse across two stainless steel pins placed between toes 1-2 and 4-5, approximately 70 mm distal to the nerve cuff implant. The hind paw was fixed in place, using a strap around the ankle, to ensure that any twitching evoked in the paw by the stimulation pulse did not produce movement artefacts in the nerve cuff; though, because the paw is largely void of muscle tissue, twitching was expected to be negligible. The ground path of the data acquisition and filtering electronics was attached to the contralateral hind paw via a 460 kΩ resistor and steel pin in series.

### 2.2. Nerve cuff with dual ring electrode array

The electrode array was produced by *The EIT Research Group, The Department of Medical Physics and Biomedical Engineering, University College London,* using the fabrication method described in [25]. The electrode array contained two columns of 14 electrodes, each 1.1 x 0.11 mm in size and 0.25 mm centre-centre spacing, running parallel to a 0.33 x 3.63 mm reference electrode on either side. When wrapped around a nerve, each column of 14 electrodes creates a ring of electrodes spaced at approximately 26º angular offset around the nerve boundary. The two columns were spaced 1 mm apart, while each column and its adjacent reference electrode were spaced 2 mm apart, Fig 2a. All electrodes were coated with poly(3,4-ethylenedioxythiophene):p-toluene sulfonate (PEDOT-pTS) coating to reduce the electrode-tissue contact impedance.

The change in longitudinal tissue impedance, parallel to nerve fibres, during neural activity is estimated to increase with decreasing inter-electrode ring spacing in modelling by [12]. This effect may be limited, however, as noted by the authors in [12], as the inter-electrode ring spacing and distance between neighbouring node of Ranvier become comparable in length. Therefore, we increased the gap between the two electrode columns from 1 mm to 3 mm (2 mm to 4 mm centre to centre) by cutting the electrode array down the centre line, and then gluing the two separated electrode columns at 3mm spacing within a 1.2 x 6 x 13 mm (ID x OD x length) silicon tube, Fig 2b. An opening which extended down the length of one side of the cuff allowed insertion and removal of the sciatic nerve during animal experiments. The nerve cuff was soaked in PBS for one hour prior to being implanted.

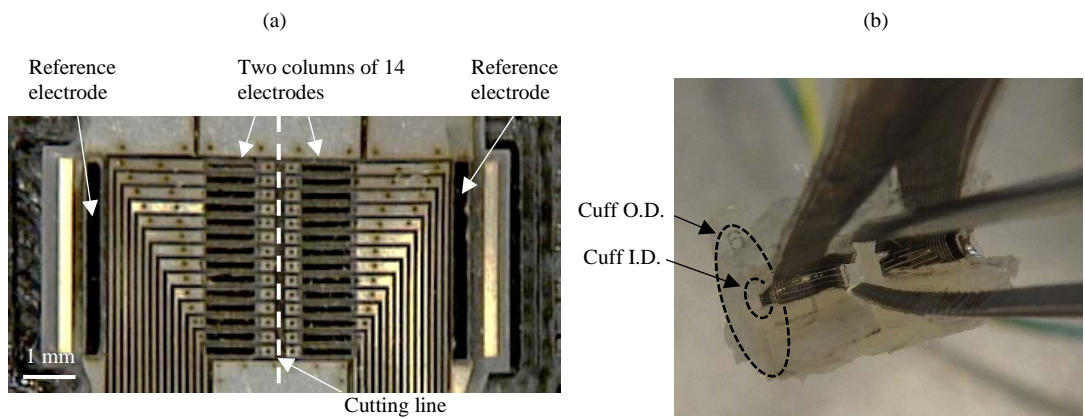

Fig 2: (a) Electrode array containing a two columns of 14 electrodes and two reference electrodes. The array was cut down the middle, along the labelled, dotted line, to increase the inter-electrode ring distance. (b) The electrode array assembled into a silicon tube to produce the dual ring nerve cuff (b) with outer (O.D.) and inner (I.D.) diameter s indicated.

### 2.3  Hardware and data collection

In single frequency and FDM experiments, one and two EIT drive currents, respectively, were applied to the nerve cuff. Each drive current was produced using a custom PCB with 6 parallel circuits each containing a waveform generator and Howland circuit current controllers, developed by *The EIT Research Group at The Department of Medical Physics and Biomedical Engineering, University College London,* and available for download from https://github.com/EIT-team. The custom PCB had 4 jumper-set current amplitudes: 16, 33, 64, 150 µA, and digitally programmable frequencies for each channel. The selected drive currents were applied continuously for the duration of each experiment.

Boundary voltage measurements were acquired from two electrodes in the nerve cuff. Each measurement signal was passed through two, in series, second order anti-aliasing filters of Sallen-Key topology, with 20 kHz cut-off frequency and $Q = 0.69$, then was logged on one of two data acquisition modules (National Instruments CompactRIO® NI9205) at 167 kS/s/ch sample rate, and 16 bit resolution across +/- 1V input range. Measurements were made with respect to the one of the two nerve cuff reference electrodes.

Data was recorded in 50 ms windows, which we may also term an 'epoch', at 2 Hz, and saved directly to .txt files for analysis in MATLAB after experiment completion. An isolated pulse stimulator (A-M Systems Model 2100) was triggered at the start of each window, via digital lines (National Instruments CompactRIO®

NI9401) and an opto-isolator, to wait 25 ms and then produce a biphasic square pulse of 200 µs duration per phase, and +/- 5 mA amplitude. Triggering and recording were synchronised using the LABVIEW FPGA interface. Each experiment ran for up to 1000 epochs, or 8 minutes 20 seconds.

The angular offset between drive electrodes for transverse drive current, across nerve fibres, was selected as 129 ° based on recommendation of 135 ° in [2], and, for longitudinal drive current selected as 0 ° and situated on different electrode-rings based on modelling in [26], Fig 3, table 1.

Two single-frequency experiments were performed initially to benchmark the extracted impedance change for comparison with FMD experiments; one in transverse direction, orthogonal to nerve fibres, and one in longitudinal direction, parallel to nerve fibres,. The drive current amplitudes were selected to minimise stimulation of fibres in the nerve cuff by the EIT drive current, as 150 µA in transverse direction [2] and 33 µA in the longitudinal direction [22]. Drive current frequencies were selected to maximise the impedance change amplitude based on results in [1], as 8 kHz for transverse and 4 kHz for longitudinal direction, table 1.

In FDM experiments, the longitudinal drive currents were selected as 33 µA such that the maximum in the multiplexed currents was in the range 50 to 60 µA. This summative range, of 50 to 60 µA, was assumed to produce minimal nerve fibre excitation based on equivalent charge per phase in the FDM-EIT drive current as that in [22], from 30 to 40 µA, 100 µs, monophasic square pulse. Minimum spacing of the drive current frequencies was selected, as 2 kHz, by Fourier transforming the demodulated signal from the initial, single frequency experiments, then doubling the main frequency band to ensure no overlap between impedance change sidebands on neighbouring carrier signals, Fig 1. Frequency spacing of 4 and 8 kHz were also implemented for comparison.

Table 1: Drive current settings and electrode positions used in each experiment

| Experiment | Number and direction of drive currents | Amplitude | Frequency | Drive and measurement electrode position |
|---|---|---|---|---|
| 1 | 1: Transverse | 150 µA | 8 kHz | drive: 0 ° / 129 ° <br> measurement: 77 ° / 205 ° |
| 2 | 1: Longitudinal | 33 µA | 4 kHz | drive: 0 ° / 0 ° <br> measurement: 51 ° / 129 ° |
| 3 | 2: Longitudinal | 33 µA | 4 kHz | drive: 0 ° / 0 ° |
|   |   | 33 µA | 6 kHz | drive: 180 ° / 180 ° <br> measurement: 26 ° / 129 ° |
| 4 | 2: Longitudinal | 33 µA | 4 kHz | drive: 0 ° / 0 ° |
|   |   | 33 µA | 8 kHz | drive: 180 ° / 180 ° <br> measurement: 26 ° / 129 ° |
| 5 | 2: Longitudinal | 33 µA | 4 kHz | drive: 0 ° / 0 ° |
|   |   | 33 µA | 12 kHz | drive: 180 ° / 180 ° <br> measurement: 26 ° / 129 ° |

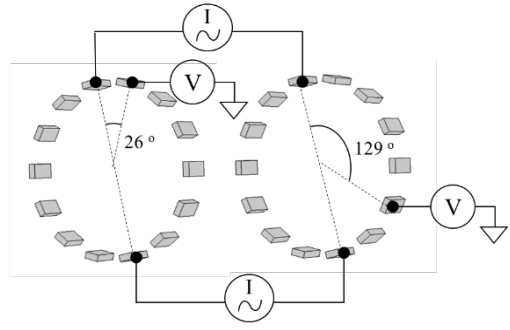

Fig 3: Drive and measurement electrode configuration for longitudinal current in the dual ring nerve cuff, used in Experiment 3, where I indicates the drive current sources electrode pairs – located at 0 ° / 0 ° and 180 ° / 180 °, and V indicates the boundary voltage measurement electrodes – located at 26 ° and 129 °. Voltage measurements are made with respect to one of the reference electrodes, which are not shown here.

*2.4.    Data processing to extract impedance changes*

The CAP was extracted from each epoch using a 6$^{th}$ order, zero phase shift, Butterworth BPF with 0.1 to 1.5 kHz pass band [1]. Between 400 and 500 epochs were then averaged together to reduce noise. The CAP frequency components were obtained via discrete Fourier transform of a 10 ms length window which commenced in the interlude between the stimulation pulse and the CAP, at 26.5 ms, and were later used in design of the BPF in the BPF-HT demodulating function used in FDM experiments.

Three pre-processing steps were implemented prior to BPF-HT demodulation: the two measurement channels in each epoch were subtracted from one another to reduce the amplitude of the stimulus artefact by removing common components; the DC component of each epoch was removed by subtraction of the mean; and, window artefacts, such as spectral leakage from transients at the window edge and Gibbs phenomena from sharp steps at the window edge, were reduced by applying a tapered window. We selected a Tukey window for this third step as the transient features in the signal are not distorted within the flat central section. The tapered sections of the Tukey window were designed to occupy 25 to 40 % of the window as this was found, using trial and error, to offer a good balance between removing ringing artefacts and maintaining an undistorted section of window large enough for analysis.

In the BPF-HT method, the BPF was implemented as a zero phase-shift Butterworth, with infinite (IIR), instead of finite (FIR), impulse response for its lower computational demand. In the two single frequency experiments, the BPF bandwidth was set to +/-2 kHz [16], and BPF order set to 6 based on an order of 5 in [2, 15, 19]. Each filtered epoch was Hilbert transformed to demodulate the transient changes in boundary voltage caused by the impedance change, then 400 to 500 epochs were ensemble averaged to reduce noise. The frequency components of the impedance change were obtained via discrete Fourier transform of a 10 ms length window which, as before, commenced at 26.5 ms, and were later used in design of the BPF in the BPF-HT demodulating function used in FDM experiments.

Tissue impedance changes $z_\%(t)$, expressed as percentage change in the demodulated boundary voltage signal $V(t)$ from the baseline value $V_i$, were calculated using equation 3:

$$z_\%(t) = 100 \left( \frac{V(t) - V_i}{V_i} \right) \qquad (3)$$

where $V_i$ was calculated as the average value across a 10 to 20 ms window of $V(t)$, corresponding to a period of largely inactive nerve tissue prior to the stimulation pulse at 25 ms.

The BPF in the BPF-HT function was designed such that the bandwidth encompassed at least 99.99% of the frequency power of the impedance change, determined from the area under the frequency-power plot. The frequency components of the CAP and impedance changes, extracted from single frequency experiments, were applied to an ideal, 100 mV amplitude carrier signal at 4 kHz in MATLAB, normalised against the maximum amplitude, and plotted next to ideal, 100 mV amplitude, unmodulated carrier signals at 6 and 8 kHz, representing frequency spacing of 2 and 4 kHz respectively. Ideal signals were used to ensure narrow spectra, and comprised exact integer division of frequency and window size with sample rate. The BPF order was selected by ensuring the intersection between the BPF response and the neighbouring carrier signal was sufficiently below the peak of the impedance change sideband, which we defined using a threshold at the frequency corresponding to the 99.99% area point on the frequency-power plot as it accounts for the size and shape of the impedance change. Similarly, we verified that the CAP did not distort the extracted impedance change by determining that its frequency components was below this same threshold within the pass band of the BPF.

In FDM experiments, BPFs, with bandwidth and order selected using the method described above, were applied during BPF-HT demodulation, and were evaluated against higher and lower order BPFs within the range 4 to 20 and in increments of 4. In addition, in line with the method presented in previous study on FDM of neural activity [20], we also applied a BPF after application of a notch filter during BPF-HT demodulation. Here, we set a BPF order of 4 (with all other BPF settings as described above), and designed the notch filter as an IIR, zero-phase shift, 6$^{th}$ order stopband filter with +/- 500 Hz bandwidth centred on the neighbouring carrier signal. Filter settings which were applied to the FMD experiments were also applied to data from the single frequency experiments to examine whether the integrity of the shape and amplitude of the extracted impedance change were maintained. As with earlier, after each epoch was filtered and Hilbert transformed, 400 to 500 epochs were ensemble averaged to reduce noise.

## 3. Results

*3.1 CAP*

In all experiments, the stimulus artefact was significantly larger than the CAPs. A 1 ms delay between ringing of the stimulus artefact, produced by the BPF, and commencement of the CAPs meant the two were visually distinguishable, Fig 4a. The CAPs commenced at 27.3 ms in experiment 1, and 28 ms in experiments 2 and 3, representing a propagation time from the centre of the stimulation pulse of 2.1 to 2.8 ms. The CAP peak-to-peak amplitudes ranged from 200 µV in experiment 2 up to 340 µV experiment 1, and the durations ranged from approximately 2.5 ms in experiments 1 and 3, to 3 ms in experiments 2. The frequency components of CAPs in all experiments showed good agreement, with the main components present between 200 to 500 Hz. A significant reduction in power was observed for all CAPs between 500 Hz and 1 kHz, a small to negligible portion

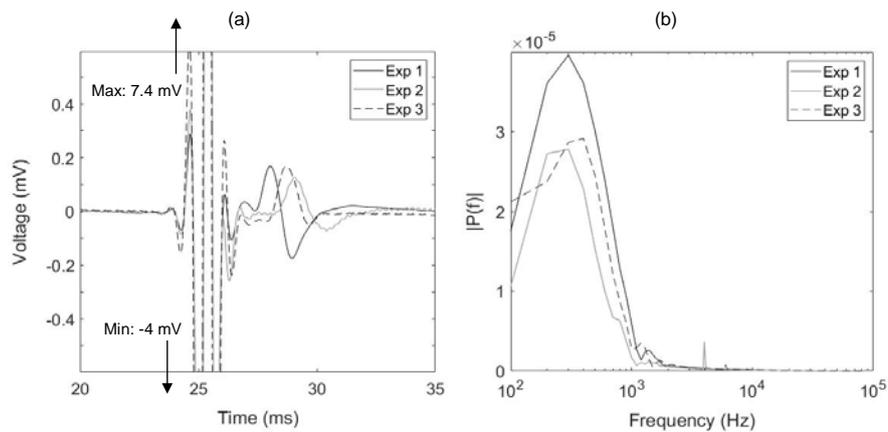

Fig 4: CAP recordings for each experiment extracted with a BPF (6$^{th}$ order 0.1 to 1.5 kHz) (a). The stimulus artefact, with amplitude peaks cropped off but indicated by labels, is visible from 23.8 to 26.6 ms, while the CAPs are visible from 27.3 to 31 ms. Frequency components of the CAPs (b) obtained via Fourier transform of a 10 ms length window of the CAP recordings.

of which may be attributable to the BPF with -3dB cut-off of 1.5 kHz.

*3.2 Pre-processing*

Subtraction of the two measurement channels significantly reduced the amplitude of the stimulus artefact, Fig 5a and b. Ringing from Gibbs phenomena was observed in ensemble averaged signals when no Tukey window was applied as a pre-processing step, Fig 5c. This ringing decayed rapidly in amplitude from a maximum at the window edge towards the centre, though with a still significant amplitude in the central 10 ms of the window relative to the amplitude of the stimulus artefact, observed at 25 ms, and the impedance change, observed from 28 to 31 ms. Ringing was up to 3 orders of magnitude lower when a Tukey window, with tapered sections occupying a total of 25 % of the window size, was applied as a pre-processing step prior to the BPF-HT function.

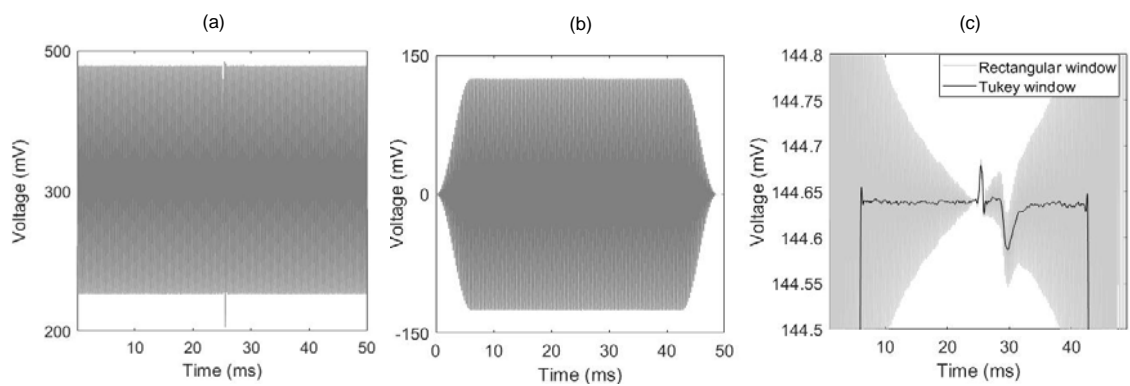

Fig 5: Boundary voltage recording from experiment 2 before (a) and after (b) pre-processing steps, with the stimulus artefact at 25 ms clearly visible in the former but not the latter, and the DC offset removed and Tukey window applied in the latter. (c) Comparison of the signal after BPF-HT (6$^{th}$ order +/- 2 kHz) and ensemble averaging with and without use of the Tukey window in pre-processing, showing that the significant ringing, which is present when no tapered window is used, distorts the impedance change which is visible between 28 and 31 ms.

*3.3 Impedance changes*

In the two single frequency experiments, the impedance change was below the noise floor (+/-0.0033% RMS) in transverse current direction, experiment 1, but clearly visible above the noise floor (+/-0.0017% RMS) in longitudinal current direction, experiment 2, Fig 6a and b. In the latter, the impedance change rose sharply at -0.0275 %/ms to a peak of -0.0358 % at 29.8 ms, then decayed back towards 0 % at a more gradual rate of between 0.016 and 0.003 %/ms. Fourier transform the 10 ms length window, containing the observed impedance change, showed main frequency components in the 100 to 200 Hz band, and significant drop in power between 200 Hz and 1 kHz.

Integration to obtain the area under the frequency-power curve, Fig 6c, revealed 99.99% of power in the 100 to 750 Hz range, Fig 6c inset. This evaluation did not consider frequencies in the 0 to 100 Hz range due to limitations of the Fourier transform imposed by the 10 ms length window used to obtain the frequencies. Therefore, a bandwidth of 750 Hz was selected for the BPF.

In determining the order of the BPF, the +/- 750 Hz points on the 4 kHz ideal carrier signal, which had been modulated with the impedance change from experiment two, established the intersection threshold at $4.5 \times 10^{-6}$ normalised amplitude, Fig 7a and b. BPF orders of 8 and 12 intersected the 6 kHz ideal carrier, and BPF order of 8 intersected the 8 kHz ideal carrier, above the threshold of the impedance change. Therefore, for 2 kHz spacing between carrier signals BPF order of 16 or higher were deemed as suitable, and for 4 kHz spacing an order of 12 or higher – or an order of 10 or higher using visual interpolation - was deemed suitable. The CAP was below the threshold within the band pass of the BPF.

In all FDM experiments the stimulation artefact was visible in the demodulated carrier signals, but was significantly smaller amplitude in the 4 kHz signal of experiments 3 and 4, Fig 8a and c. The impedance change was clearly visible in all demodulated carrier signals between 27 and 31 ms, but appeared distorted in the 4 kHz signal of experiment 3 due to a -0.01 % drop in impedance across the artefact Fig 8a, possibly caused by noise or other electrical disturbance. The amplitude of the impedance change varied between frequencies within the same experiment, and across experiments for the same frequency Fig 8a-f.

Ringing introduced by the neighbouring signal was present in signal demodulated with an $8^{th}$ order BPF in experiment 3, Fig 8a-b, $4^{th}$ order BPF in experiment 4, Fig 8c-d, and $4^{th}$ order BPF of the 4 kHz signal in experiment 5, Fig 8e. Ringing from Gibbs phenomena, caused by the relatively large step in the Tukey window, was present in all demodulated signals, and, as expected, exhibiting highest amplitude at the edges of the flat section of the Tukey window, at 11 and 39 ms, and increasing amplitude for higher order BPFs. Distortion of the impedance change introduced by BPF of the neighbouring stimulus artefact, observed as variations in the shape and amplitude of the impedance change, was present in all signals, but particularly pronounced in the 6 kHz signal in experiment 3, Fig 8b, and 12 kHz signal in experiment 5, Fig 8f.

Filtering with a $4^{th}$ order BPF and the notch filter removed ringing introduced by the neighbouring signal in experiments 4 and 5, Fig 8c-f, and significantly reduced this ringing in experiment 3, Fig 8a-b. A $6^{th}$ order BPF and the notch filter removed this ringing in experiment 1.

Comparison of the 8, 12, 16 and 20 order BPFs to the $6^{th}$ order BPF in the 4 kHz single frequency experiment, experiment 2, showed the higher order BPFs introduced more ringing from Gibbs phenomena, which distorted the impedance change peak amplitude by up to -0.0001% (from -0.003612 % to -0.003622 %), Fig 8g.

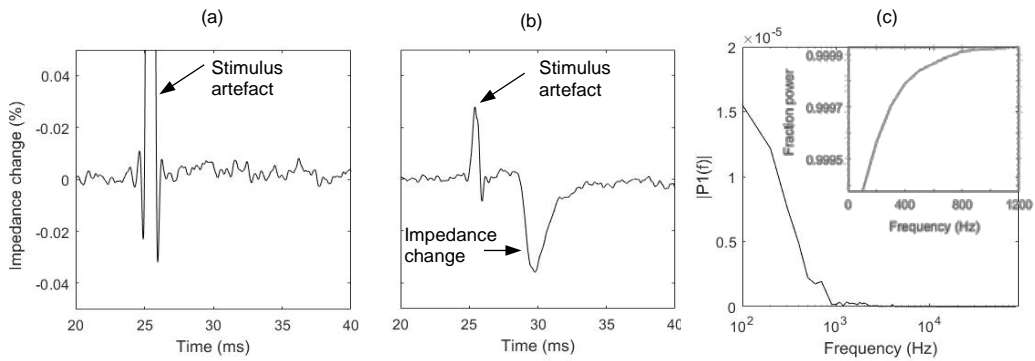

Fig 6: Impedance changes extracted from single frequency experiments in transverse (a) and longitudinal (b) current directions, expressed as percentage change from baseline values. The stimulus artefact is visible at 25 ms in both (a) and (b), and an impedance change is visible in (b) at 28 to 32 ms. Frequency components of the impedance change (c) are dominant below 1 kHz, with 99.99% of the power contained with the 100 to 750 Hz band (c inset).

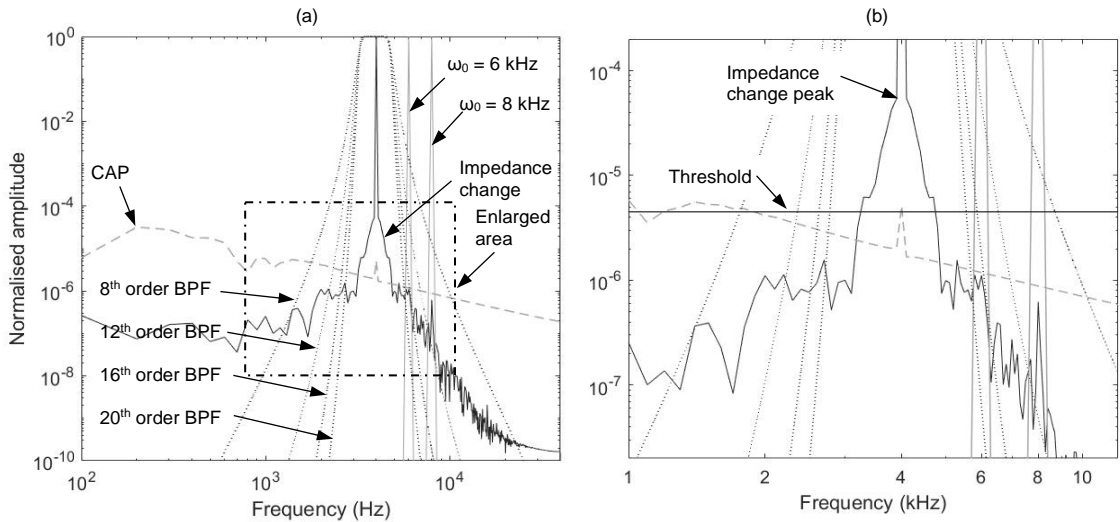

Fig 7: Frequency components of the impedance change from experiment 2, modulating an ideal 4 kHz carrier signal, and the CAP from experiment 2, plotted next to two ideal carriers, one at 6 kHz and one at 8 kHz. Overlaid are the responses of four Butterworth IIR BPF filters with bandwidths of +/-750 Hz and orders of 8, 12, 16 and 20 (a), and, in the enlarged area of (a), close-up of the impedance change and surrounding features (b), showing the impedance peak, and the threshold used to determine suitable BPF order.

## 4. Discussion

The CAP propagation time, duration, and frequency components were in broad agreement with previous studies which stimulated the sciatic nerve using pins through the hind paw [1, 27]. The corresponding propagation velocity of 25 to 35 m/s agrees with the expected physiology of the excited fibres, which, due to the lack of muscles in the paw, are dominated by smaller diameter fibres associated with temperature and touch, and contain minimal larger diameter motor and proprioception fibres.

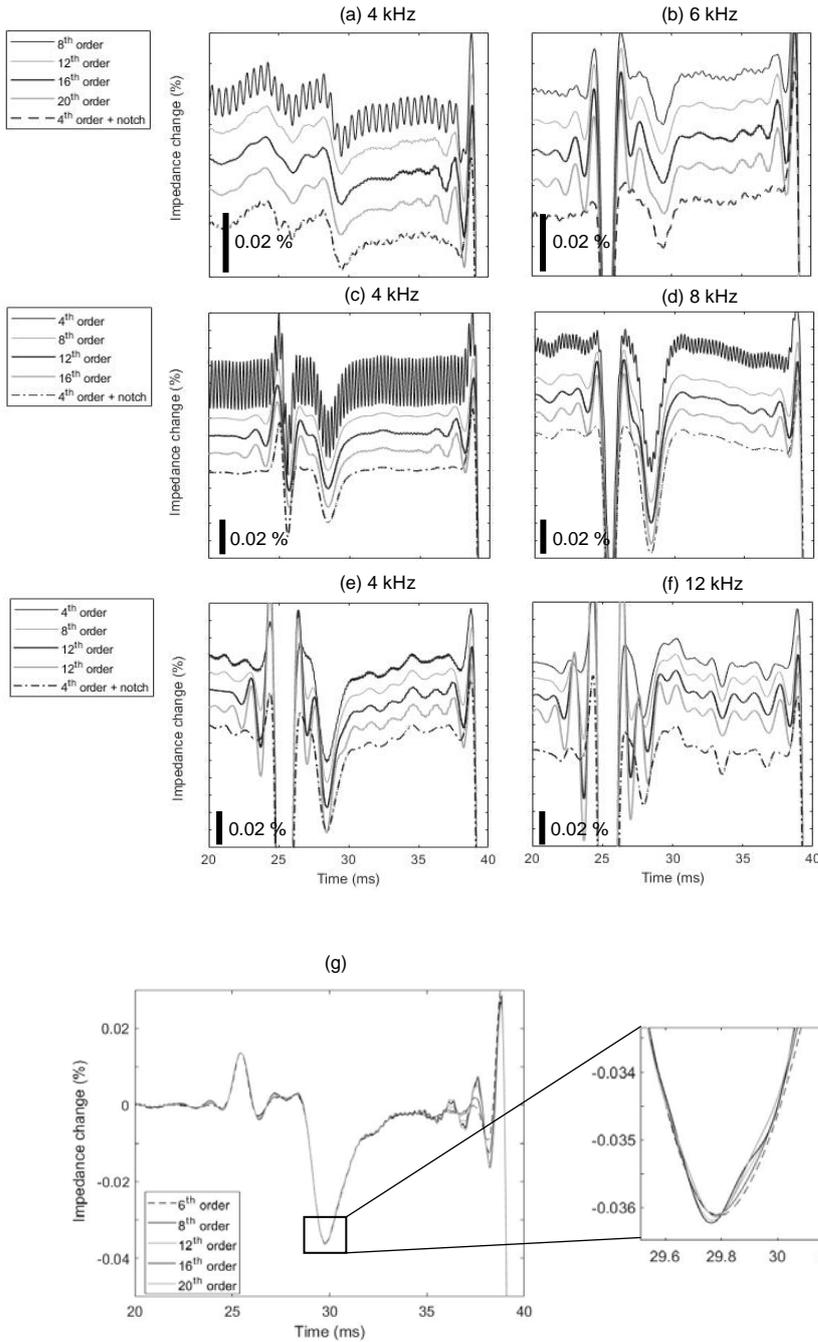

Fig 8: Demodulated impedance changes from the 4 (a) and 6 (b) kHz carrier signals in experiment 3, the 4 (c) and 8 (d) kHz carrier signals in experiment 4, the 4 (e) and 12 (f) kHz carrier signals in experiment 5. Ringing at the window edge is larger at higher BPF orders (a-g). Residual ringing from the neighbouring carrier signal is visible at lower BPF orders (a-d), but is reduced significantly by a notch filter in (a-d). Higher order BPFs distort the impedance change (g).

Subtraction of the two measurement channels reduced the stimulus artefact, and is consistent with practical application of EIT where differential measurement between electrode pairs is common [26, 28].

The Tukey window significantly reduced ringing from Gibbs phenomena at the window edge compared to rectangular window, Fig 5c, though, as expected, ringing was still present at the edge of the flat section of the Tukey window at higher BPF orders, Fig 8a-f. With Gibbs phenomena, the decay in ringing towards the centre of the window favours larger temporal windows, such as the 200 ms and 500 ms windows implemented in [2] and

[19], respectively. However, windowing introduces a delay between occurrence of the neural event and its extraction, which, in the context of neural prosthetics, has direct influence on user experience through intentional binding effects [29], and practical repercussions on prosthetics control and user acceptance rates [30].

In single frequency experiments, the unobserved impedance change with transverse current may be due to the low resolution in the data acquisition module, though this limitation is offset by the dithering effect of significant system noise, circa +/- 170 µV RMS, and ensemble averaging. The location of the impedance change with longitudinal current, experiment 2, was consistent with that of the corresponding CAP, Figs 4a and 6b. The frequency components of the impedance change were comparable to those of the CAP, where the former was dominant between 100 to 200 Hz and limited to below 1 kHz and the latter dominant between 200 and 500 Hz and limited to below 1 kHz, Fig 4b and 6c. The difference between these dominant frequency bands, by a factor of approximately 2, may be due to the biphasic-pulse shape of the CAP versus the monophasic pulse-shape of the impedance change Figs 4a and 6b.

Variation in the amplitude of the impedance change across experiments for the same 4 kHz carrier signal frequency are caused by differences in tissue morphology of the animals, electrode positions relative to this morphology, and electrode impedance. Variation in the amplitude of the impedance change across different frequencies in the same experiment cannot be compared to previous modelling in [1, 12] or results in [22] as the angle between drive and measurement electrodes were different for each FDM carrier signal.

The selected BPF bandwidth, +/-750 Hz, is lower than that used in previous EIT studies on mammalian peripheral nerve, +/- 2 [16] to +/- 3 kHz [2]. This difference in bandwidth may be rationalised by differences in the stimulation method, via the paw in the current study and via the tibial and peroneal nerve branches in [2, 16], where, in the latter, the stimulation of all fibre diameters and less dispersion from a shorter propagation distance to the recording nerve cuff would both produce a shorter duration, higher frequency CAP and impedance change. Other considerations when selecting the BPF bandwidth are its inverse relationship with the temporal resolution, and proportional relationship with noise, in the demodulated signal.

The BPF orders indicated by filter modelling to avoid ringing from the neighbouring signal, Fig 7, of 16th or higher and 10th or higher with 2 and 4 kHz frequency spacing respectively, were both significantly higher than those used in previous single frequency EIT studies on neural activity, 5th order in [2, 15, 19]. FMD experiments showed that filter modelling tended to overestimate the BPF order required, with 12th order and 8th order proving sufficient for 2 and 4 kHz carrier spacing, respectively, Fig 8a-d. While the distortion introduced by higher order BPFs to the impedance change in experiment 2 was relatively minor, Fig 8g, higher frequency and larger amplitude impedance changes would generate more distortion, potentially producing imaging artefacts and errors when applied to EIT reconstruction.

When the BPF was used in conjunction with the notch filter, residual ringing from the neighbouring signal was either removed or significantly reduced, and, due to the low BPF order, the least ringing from Gibbs phenomena of the orders evaluated. These results indicate that the BPF + notch filter implementation presented in [20] provides better performance than BPF alone due to the lower possible BPF orders. Notch filter settings were not examined as part of this current study, but should also be considered as they introduce another potential source of ringing and distortion.

**5. Conclusion**

We have characterised the frequency components of the CAP and impedance change which are produced with, and are specific to, paw stimulation and sciatic nerve recording in rat. We then demonstrated a framework, using these frequency components, for selecting band pass filter settings in BPF-HT demodulation of impedance changes. A Tukey window and notch filter and considered critical processing steps to avoid unwanted ringing artefacts in the demodulated signal.

**Acknowledgements**

Authors would like to thank David Holder, who leads The EIT Research Group at University College London, for providing essential hardware in support of this research; Marshall Lim from The Department of Mechanical Engineering, The University of Auckland for his help with development of the system software and hardware; Zaid Aqrawe and Darren Svirskis from the School of Pharmacy, The University of Auckland for their help with PEDOT coating electrodes; and staff at the School of Biological Sciences, The University of Auckland for their help with animal handling.